\documentclass[conference]{IEEEtran}
\IEEEoverridecommandlockouts
\usepackage{graphicx}

\graphicspath{ {./Pics/} }
\usepackage{amsmath}

\usepackage{tablefootnote}

\usepackage{cite}

\ifCLASSINFOpdf
\else
\fi
\usepackage{array}
\usepackage{url}

\usepackage{stackengine}

\usepackage{multicol}

\usepackage{xcolor}
\usepackage{enumitem}
\usepackage{float}

\usepackage{flushend}
\begin{document}
%
\title{A Bayesian Approach to Probabilistic Solar Irradiance Forecasting}

\author{\IEEEauthorblockN{Kwasi Opoku, Svetlana Lucemo, Qun Zhou Sun, Aleksandar Dimitrovski}
\IEEEauthorblockA{\textit{Department of Electrical \& Computer Engineering} \\
\textit{University of Central Florida}\\
Orlando, Florida 32826\\
Email: opokukwasi@knights.ucf.edu, sdlucemo@knights.ucf.edu, QZ.sun@ucf.edu, adimitrovski@ucf.edu}

}


%
 
\makeatletter
\def\ps@IEEEtitlepagestyle{%
  \def\@oddfoot{\mycopyrightnotice}%
  \def\@evenfoot{}%
}
\def\mycopyrightnotice{%
  {\footnotesize 978-1-6654-9921-7/22/\textdollar 31.00 ©2022 IEEE\hfill}
  \gdef\mycopyrightnotice{} 
}

\maketitle

\begin{abstract}
The output of solar power generation is significantly dependent on the available solar radiation. Thus, with the proliferation of PV generation in the modern power grid, forecasting of solar irradiance is vital for proper operation of the grid. To achieve an improved accuracy in prediction performance, this paper discusses a Bayesian treatment of probabilistic forecasting. The approach is demonstrated using publicly available data obtained from the Florida Automated Weather Network (FAWN). The algorithm is developed in Python and the results are compared with point forecasts, other probabilistic methods and actual field results obtained for the period.
\end{abstract}

\begin{IEEEkeywords}
Solar irradiance, Bayesian method, probabilistic forecasting, quantile regression
\end{IEEEkeywords}

%
\IEEEpeerreviewmaketitle

\section{Introduction}
The new challenges to operation of power systems with high penetration of renewable resources have been extensively discussed in various literature \cite{kim2010new, 5666114,luo2015overview}. Prominent among these is the intermittent output of such distributed generation. In the case of conventional sources, the unit commitment of  power generation is dependent largely on forecasts based on variation in demand. However, particularly for PV sources, the intermittency in output is also influenced majorly by the changes in the solar irradiance, which is superimposed on the demand response\cite{murata2018modeling}. In order to mitigate the effects of this challenge, researchers and system operators have placed significant focus on improving the quality of solar irradiance forecasts.

Approaches to forecasting solar irradiance may be categorized based on the forecast methods used, whether physical or non-physical \cite{antonanzas2016review}. Physical methods involve detailed modelling of the relationship between predicted values and the natural observations (satellite observations, Numerical Weather Predictions (NWP), etc). Such physical modelling tend to lead to intensive computation with complex numerical techniques\cite{li2020review}. Non-physical methods, on the other hand, depend on historical weather data without detailed physical models of relationships between inputs and output. Thus, they are statistical. 

Early statistical methods for forecasting considered the properties of the aggregated historical data used \cite{goh1977stochastic}. Many of the commonest forecast methods are based on such regressive processes. These include the autoregressive, autoregressive moving average (ARMA) and the autoregressive integrated moving average (ARIMA) methods\cite{aguiar1992tag, craggs2000statistical}. A major drawback to these methods is that the statistical transformations applied to the climatic data used for solar irradiance prediction may adversely impact the accuracy of the results \cite{ahmad2015hourly}.

Artificial Intelligence methods have been proposed by many researchers as an alternative to improve the earlier statistical approaches. Particularly, Artificial Neural Network (ANN) techniques have been explored and implemented to improve the accuracy of the statistical methods. ANNs create a non-linear mapping between input data and output variables. This makes the technique desirable for applications involving time series of weather data\cite{voyant2017machine}. It is shown in \cite{mellit2008artificial} that the overwhelming majority of ANN methods used for solar irradiation forecasting use a multilayer perceptron (MLP). Other neural network (NN) methods are prosposed in existing literature, including Time Delay Neural Network (TDNN), Convolution Neural Netowrk (CNN), among others. A comprehensive review of their performance and applicability are presented in \cite{mellit2008artificial}. Generally, the adoption and scalability of many of such ANN techniques in other forecasting case studies are hindered by the fact that the parameters that determine the prediction are many and very varied \cite{voyant2017machine}.

Due to the varied behavior of atmospheric data, many researchers consider stand-alone prediction models insufficient for forecasting. This has led to the development, in recent years, of a myriad of hybrid models for solar radiation forecasting\cite{guermoui2020comprehensive}. These models combine the unique features of different single prediction models to explore different patterns in data and improve the accuracy of estimated values. For instance, approach used in \cite{cao2005forecast}, combined ANN with a feature tool, wavelet analysis. Similarly, in \cite{reikard2009predicting,ji2011prediction}, methods such as ARMA and ARIMA were used together with TDNN.

Generally, the performance of single value predictions or point forecasts can be assessed by using a cost function of the errors in the prediction. \cite{fatemi2018parametric}. Such performance evaluation methods include Root-Mean-Squared Error (RMSE), Mean Absolute Percentage Error (MAPE) and Mean Absolute Error (MAE). The RMSE is used in Section IV in our case study to evaluate the point forecast, and it uses a Eucledian distance to represent the distance of the forecast point from the measured value. Thus, a smaller value represents a more accurate prediction. It is evaluated using the expression;

\begin{equation}
RMSE = \sqrt{\frac{\sum_{i=1}^{N}||y(i)-y_*(i)||^2}{N}}
\label{RMSE}
\end{equation}
where \(N\) is the number of data points, and \(y(i)\) and \(y_*(i)\) represent the ith actual measured value and ith prediction respectively. Developers are able to tune the hyperparameters affecting the accuracy of the prediction based on the errors evaluated. 

A major challenge that has been identified, however, with many forecasting techniques is the uncertainty that is associated with single value predictions. The impact of over- and under-forecasting for a system with large-scale penetration was carried out in \cite{martinez2016value}. It was observed that the inclusion of a variable for uncertainty in solar forecasts reduced overall operational costs, by reducing fuel costs and start and shutdown costs for conventional thermal plants. Thus, confidence intervals and probabilistic forecasting techniques are currently used to enhance the operational benefits of solar forecasting\cite{fatemi2018parametric, join2016solar}. These methods provide a risk-based decision-making in system operation by providing a more complete stochastic characterization of the input data and estimated values obtained.

This paper, therefore, discusses a Bayesian treatment of probabilistic forecasting. By employing the Bayesian principle, the prior and posterior uncertainties in forecasting data, are properly modelled to provide a more informed prediction. The algorithm developed is evaluated using weather data obtained from the Florida Automated Weather Network (FAWN). The remainder of the paper is organized as follows. Section II discusses probabilistic forecasting whereas the methodology for applying Bayesian approach to probabilistic forecasting is discussed in Section III. The performance of the method is evaluated with a case study in Section IV and Section V concludes the paper.

\section{Probabilistic Forecasting}
Probabilistic forecasting methods are commonly employed as a more applicable approach to solar forecasting. For operational planning purposes, short-time forecasts of conditions such as solar irradiance are vital. Resulting time-series predictions, thus give way to fluctuations and make single point forecasts inherently imperfect. Probabilistic methods are thus used to explicitly capture the uncertainties associated with the predictions, often by using probability distributions.

\subsection{Types of Probabilistic Forecasting Methods}
Probabilistic forecasts may be considered to be parametric or non-parametric. In parametric methods, the forecasted value is assumed to follow a known prior distribution\cite{li2020review}. For instance, a point forecast is combined with a distribution of it's forecasting errors, which gives the operator an idea of the possible accuracy based on a prior distribution \cite{david2016probabilistic}, like a Gaussian. Parametric methods are simple. However more accurate results may generally be obtained from such methods based on the symmetry between the distribution assumed and the actual prior data.

Non parametric methods generally provide an improved performance over the parametric methods by attempting to match the asymmetry in the prior observed data. Quantile regression is a common non-parametric method\cite{lauret2017probabilistic}. It calculates different percentages (quantiles) of the target values. This is achieved by minimizing the loss function for different quantiles \cite{panamtash2020copula}. Various ensemble forecast methods also use non-parametric methods. These methods combine different forecast methods to collectively produce an improved forecast that is more accurate than the singular predictions of the individual methods involved \cite{sperati2016application}.

\subsection{Performance Evaluation}
Methods such as RMSE and MAE used for evaluating the performance of point forecasts cannot be used to evaluate probabilistic forecasts. Scoring rules are used in this case by assigning a numerical score based on the distribution predicted and the actual realized output\cite{hersbach2000decomposition}. A summary of some of the common scoring rules and evaluation metrics for probabilistic forecasting are presented below.

The Logarithmic Score \(LogS\) compares the predicted probability distribution function (PDF) with the actual realized output using (\ref{LogS})\cite{fatemi2018parametric}.

\begin{equation}
LogS(f,x) = -log(f(x))
\label{LogS}
\end{equation}
where \(f\) is the predicted PDF. Since it captures the negative logarithm of the likelihood function, a lower score indicates a better performance of 
the prediction.

One of the most commonly used scoring rule for probabilistic forecasting is the Continuous Ranked Probability Score (CRPS). It provides an idea of a distance between the forecast and the actual realized value, \(x_{a}\). For a cumulative distribution function, \(F\), as shown in \cite{hersbach2000decomposition}, the CRPS is given by

\begin{equation}
CRPS(F,x_{a}) = \int_{-\infty}^{\infty} [F(x) - F_{a}(x)]^{2}dx
\label{CRPS1}
\end{equation}
where \(F_{a}(x)\) is given by

\begin{equation}
F_{a}(x) = H(x - x_{a})
\label{CRPS2}
\end{equation}
and \(H\) is the Heaveside function given by

\begin{equation}
H(x) = \begin{cases}
1 & \text{if } x \geq 1\\ 
0 & \text{if } x <  0 
\end{cases}
\label{CRPS3}
\end{equation}
From (\ref{CRPS1}), it can be seen that the CRPS integrates the square of the differences between the predicted CDF and the Heaveside function (a CDF that is either 0 at any value less than the actual, or 1 at the actual value). 

For quantile forecasts, the Pinball Loss is often used as a performance evaluation metric. For an observation \(x\) made, and a forecast \(\hat{x}_{q}\) for the quantile \(q\)\, the pinball Loss, \(L_{q}\) is given by: \cite{van2018review}

\begin{equation}
L_{q}(\hat{x}_{q},x) = \begin{cases}
(1-q)|\hat{x}_{q}-x| & \text{if } \hat{x}_{q} \leq x\\ 
q|\hat{x}_{q}-x| & \text{if } \hat{x}_{q} >  x 
\end{cases}
\label{pinball}
\end{equation}
An average across all data points in various quantiles is used to provide a score that indicates bandwidth of the prediction interval. The smaller the value therefore, the more accurate the prediction. 

Several other evaluation metrics have been proposed including graphical methods such as the Reliability Diagram and the Rank Histogram \cite{david2016probabilistic}. In this paper, the CRPS is used to evaluate the performance of the forecast technique.

\section{Application of Bayesian Approach to Solar Forecasting}

\subsection{Bayesian Method}

The Bayesian approach is used in this paper to model the uncertainties in the parameter estimates. Consider a set of input data, \(x = \left \{x_{1},...,x_{n}  \right \}\), and set of target outputs, \(y = \left \{y_{1},...,y_{n}  \right \}\). From simple linear regression, the output \(y\) is given by

\begin{equation}
y = \omega_{0} + \omega_{1}x_{1} + \omega_{2}x_{2} + ... + \omega_{m}x_{m}
\label{linreg1}
\end{equation}
where \(\omega\) represents the set of parameters or weights used to determine the output \(y\). Equation (\ref{linreg1}) could be re-written as

\begin{equation}
y = x^{\top }\omega
\label{linreg2}
\end{equation}

A Bayesian treatment allows for a probability distribution to be applied to both the parameters, \(\omega\), and the predictions, \(y\), and thus it can capture the uncertainties that may be associated with both sets of values. From Baye's Theorem, the probability of the parameters \(\omega\), given target values \(y\) and the input data \(x\), can be evaluated from the expression

\begin{equation}
p(\omega|x,y)  = \frac{p(y|x,\omega)p(\omega)}{p(x,y)}
\label{bayes}
\end{equation}
where \(p(y|x,\omega)\) is the Likelihood, ie probability of the target values given the input data and parameters; \(p(\omega)\) is the Prior probability, ie the initial knowledge of the probability of the parameters; \(p(x,y)\) is the joint probability of the input data and the targets. $p(\omega|x,y)$ is the Posterior or probability of the parameters given the input data and targets. For this paper \(X\) and \(Y\) will be used to represent the set of input data and targets.

Thus, using Baye's theorem an initial distribution (Prior probability) is obtained for the parameters. The posterior distribution of the parameters obtained will therefore follow a similar distribution which is is then used to  obtain a posterior prediction of the target values, producing a probabilistic prediction. 

With respect to solar output forecasting, the dependent variables, $y$, will represent the solar irradiance to be predicted. The independent variables, $x$, will be the input weather data that is trained for prediction whereas the parameters, $\omega$, will represent the dependencies between $y$ and $x$, as seen in (\ref{linreg1}). 

In recent probability and statistics models, the application of copulas have gained increasing popularity in describing the dependency structure between different continuous distributions \cite{daneshkhah2016probabilistic}. Previous knowledge of the data can be used to select a copula that properly describes the dependency between the input weather data and the output solar irradiance. Such a method of assuming and selecting a dependency structure to estimate the parameters employs Parametric Copulas. In this paper, a simple Gaussian copula (parametric) method is adopted to obtain parameters prior. Thus the marginal distribution of temperature is mapped to the normal distribution with a mean of 0 and standard deviation of 1. Thus, for instance, the 95th percentile of temperature distribution will be mapped to 95th percentile of solar irradiation distribution and will have a value of 1.645.

An alternative method to using parametric copulas is the application of empirical copulas which do not assume the dependency structure. This approach estimates the copulas from the historical data based on a kernel density estimation \cite{panamtash2020copula}. Empirical copulas generally provide more accurate representation of the dependencies and thus, lead to more accurate results but may involve complex computations.

\subsection{Prior Distribution}
Based on previous knowledge, copulas are selected for obtaining the parameters and a prior distribution can be assumed for the parameters. Let us consider a prior distribution and the model for the likelihood to be given respectively by,
\begin{equation}
p(\omega)=\mathcal{N}(\omega|\mu_{0},\sigma_{0})\
\label{prior}
\end{equation}
\begin{equation}
p(Y|X,\omega)=\mathcal{N}(Y|\Phi ^{\top }(x)\omega,s^{2})
\label{likelihood}
\end{equation}

 \(\mathcal{N}\) represents a Gaussian distribution and \(\mu_{0}\) and \(\sigma_{0}\), the prior mean and standard deviation of the distribution respectively.
 The likelihood is also assumed to be Gaussian, with \(\Phi^{\top}(x)\) representing a vector form of the dataset. \(s\) is the standard deviation.
 
Sufficient insight on the historical weather patterns allows for a more accurate and representative distribution and copula to be used as prior distribution of parameters in order to obtain an unbiased process.
If a Gaussian distribution with a mean of \(\mu_{0}\) and a standard deviation of \(\sigma_{0}\) is assumed for the parameters, as above, the expected posterior distribution of the parameters obtained will also be Gaussian.

\subsection{Posterior Distribution}
The posterior distribution of the parameters, $p(\omega|X,Y)$,  is used eventually to determine the posterior prediction of the target. Given the \(p(\omega)\) as the distribution of the copulas used in the prior distribution, represented by \(\mathcal{N}(\omega|\mu_{0},\sigma_{0})\), then $p(\omega|X,Y)$ can also be represented by a Gaussian distribution \(\mathcal{N}(\omega|\mu_{N},\sigma_{N})\), where the new mean \(\mu_{N}\) and standard deviation \(\sigma_{N}\) may be estimated in closed form as: 

\begin{equation}
\sigma_{N} = (\sigma_{0}^{-1} + s^{-2} \phi^{\top}\phi)^{-1} 
\label{posterior_mea}
\end{equation}
\begin{equation}
\mu_{N} = \sigma_{N}(\sigma_{0}^{-1}\mu_{0} + s^{-2} \phi^{\top}y) 
\label{posterior_sd}
\end{equation}
The subscript \(N\) is used to represent the size of the data set. An extensive proof of this closed form expression is provided in \cite{deisenroth2020mathematics}.

\subsection{Posterior Prediction}
Having obtained the posterior distribution of the parameters, these are then used to obtain a predictive distribution of the solar irradiance, \(y_{*}\) (probabilistic forecast) given new input weather data \(x_{*}\). Such a distribution, in this case, will be a univariate Gaussian based on the distribution adopted for the parameters. This can be obtained by

\begin{align}
p(y_{*}|X,Y,x_{*}) &= \int p(y_{*}|x_{*},\omega)p(\omega|X,Y)d\omega \nonumber \\
&=\int \mathcal{N}(y_{*}|\Phi ^{\top}\omega,s^{2}) \mathcal{N}(\omega|\mu_{N},\sigma_{N})d\omega \nonumber \\
&= \mathcal{N}[y_{*}|\Phi ^{\top}(x_{*})\mu_{N},\Phi ^{\top}(x_{*})\sigma_{N}\Phi (x_{*})+s^{2}]
\label{postpred}
\end{align}

It can be realised from (\ref{postpred}) that the predictive distribution has a mean given by (\(\Phi ^{\top}(x_{*})\mu_{N}\)) and variance (\(\Phi ^{\top}(x_{*})\sigma_{N}\Phi (x_{*})+s^{2}\)).

\section{Case Study}
\subsection{Data}
The solar radiation data used to train and test the proposed Bayesian algorithm is based on data obtained around the Orlando, Florida, area. This is publicly available data obtained from the Florida Automated Weather Network (FAWN) database \cite{fawn}. It consists of 42 weather stations and provide various atmospheric measurements including Temperature at different heights, Relative Humidity, Rainfall, Wind Speed and Solar Irradiance. It is able to generate meaurements at a minimum time resolution of 15 minute interval. This is adopted for this paper. Also for this study, forecasts of solar irradiance in the Orlando area for April 26th, 2022 are predicted. Forecasts for the times 11:45am to 2:00pm were particularly desired. The measurement data from the Apopka weather station from January 1, 2021 to April 25th, 2022 is thus used in this study as training and test data. 

\begin{figure}[ht]
\centering
\includegraphics[width=0.51\textwidth]{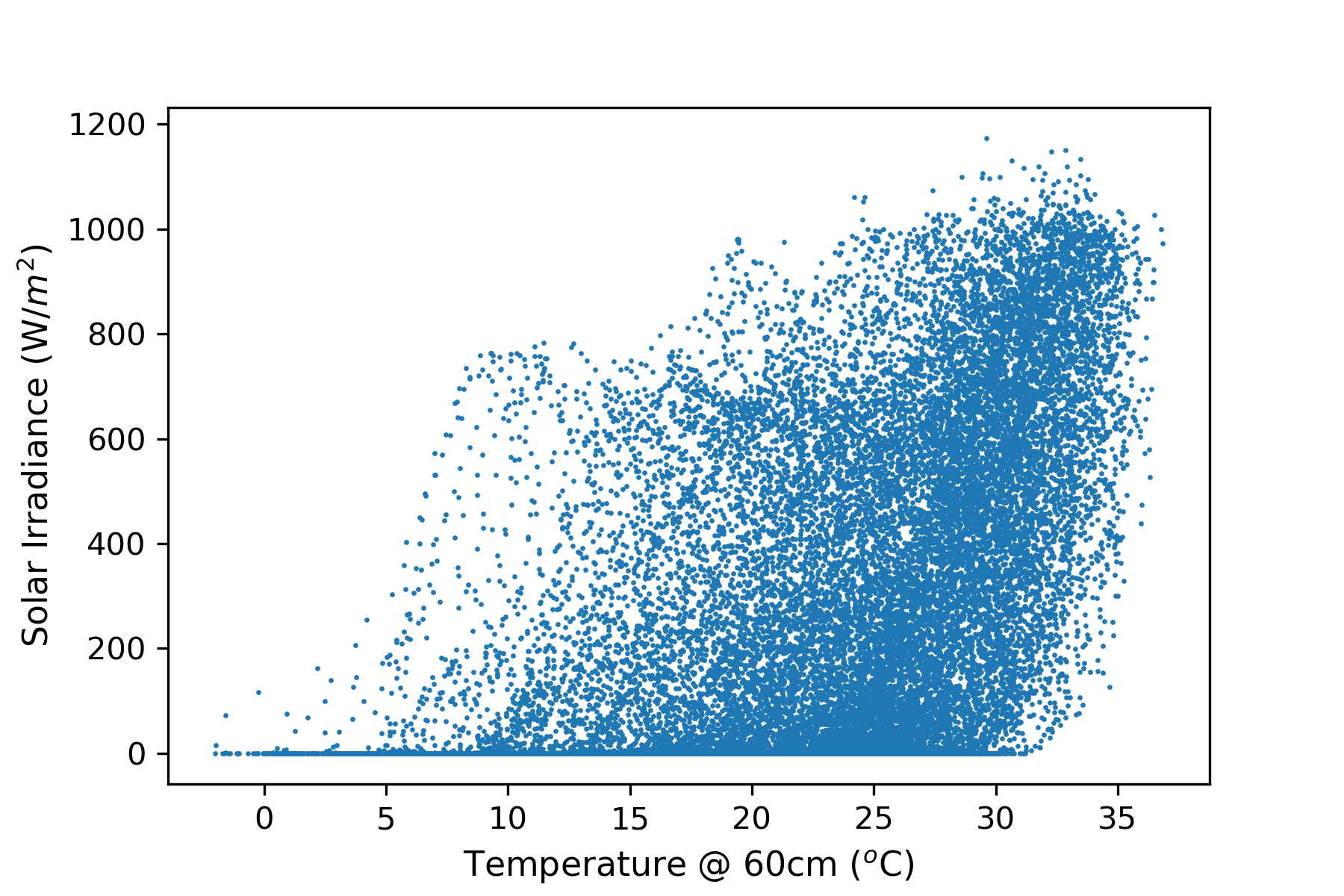}
\caption{Scatter Plot of Temperature and Solar Irradiance}
\label{scatter}
\end{figure}

Based on correlation analysis, the Temperature at 60cm (in $^o$C) revealed the strongest correlation with the solar irradiance. The target data is the Solar Irradiance (in W/m$^2$). While it is known that solar irradiance is more influenced by cloud cover than temperature, this case study attempts to use the proposed method to find a pattern between the latter and irradiance. The success of this will be benefitial in the absence of cloud cover data as in the case of the FAWN data. A scatter plot of the solar irradiance with respect to the temperature is shown in Fig. \ref{scatter}

\subsection{Point Forecasting Method}
\begin{figure}[!t]
\centering
\includegraphics[width=0.51\textwidth]{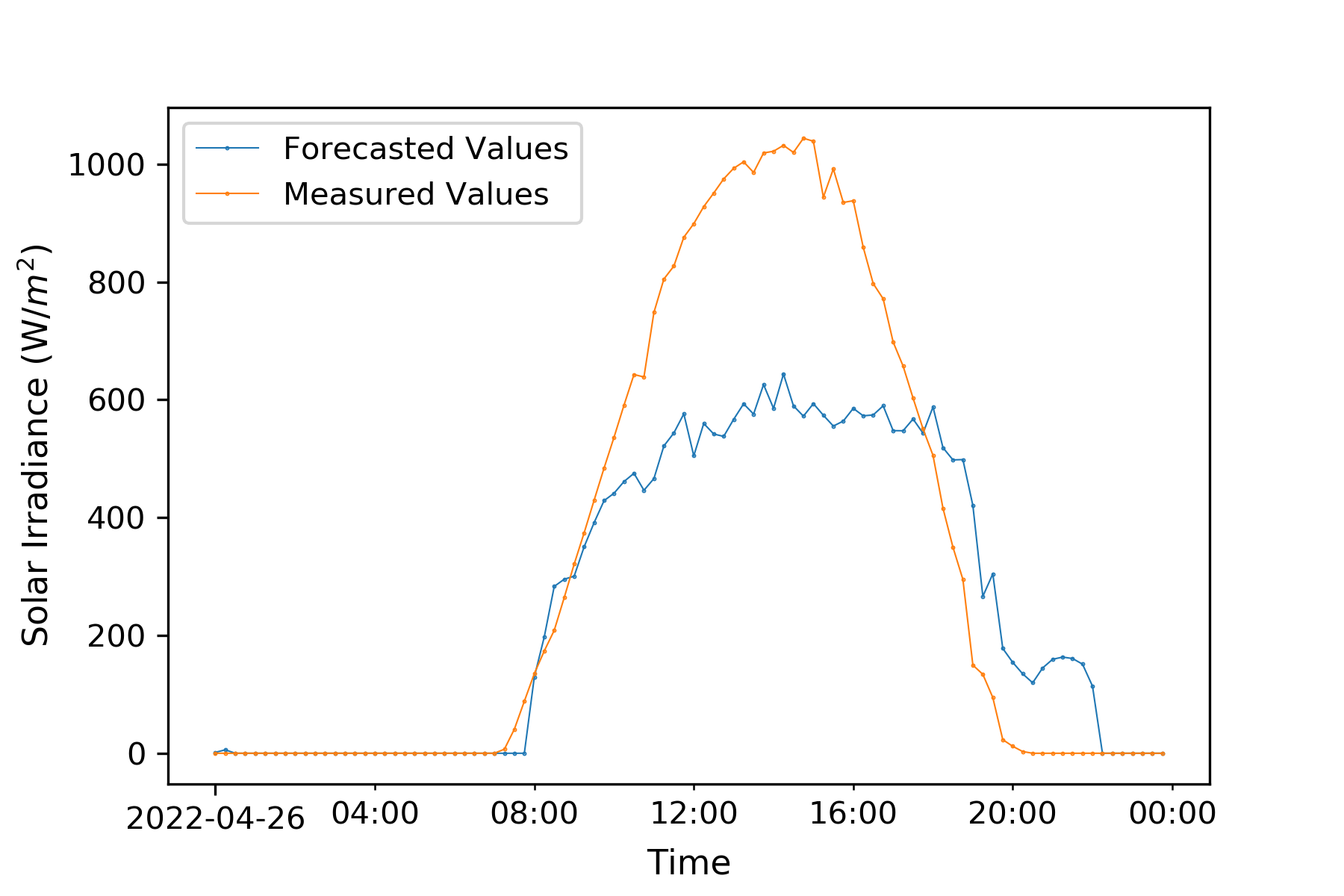}
\caption{Point forecast and Measured Solar Irradiance for 26th April}
\label{point}
\end{figure}

A multiple linear regression approach was used to obtain point forecasts for the 26th of April, 2022. Weather data including Temperatures at 2cm, 10cm and 60cm, Wind Speed and Wind Direction were selected as independent variables to predict the solar irradiance. Fig. \ref{point} shows the forecast results, compared with the actual data measured on the day.

The RMSE, discussed in Section I, is used to evaluate the performance of the model, and a value of 134.3 is obtained. This shows that the point prediction values are significantly lower at critical parts of the day, mainly in the afternoon, with peak solar irradiance output This is due to the fact that the linear regression approach provides point forecasts as the mean of the irradiance. Although, this in itself may not be unexpected of forecast methods, such a deterministic method does not provide a risk-based approach to evaluate the possible variation of the results obtained. This poses a challenge to grid operation.

It can be observed that the actual measured values for 26th April, 2022 have the semblance of the clear-sky irradiance. Thus the clear-sky model using Global Horizontal Irradiance (GHI) and Direct Normal Irradiance (DNI) data are presented for comparison. Details of these irradiance calculations are provided in \cite{ineichen2002new} and \cite{law2014direct}, but are not discussed in this paper due to space constraints.

\begin{figure}[!t]
\centering
\includegraphics[width=0.51\textwidth]{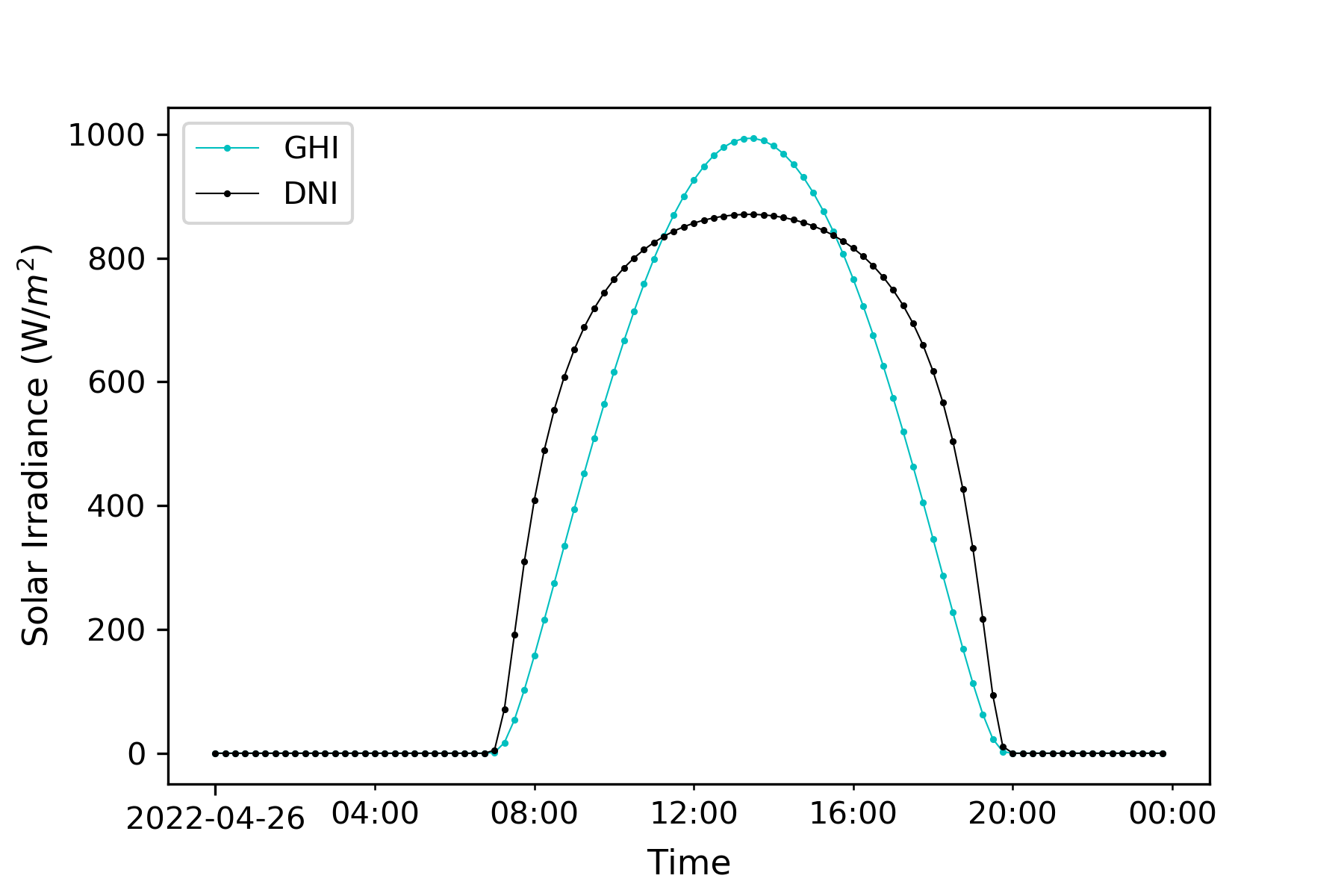}
\caption{GHI and DNI for 26th April}
\label{ghi_dni}
\end{figure}

Fig. \ref{ghi_dni} shows the GHI and DNI values. RMSE of 64.89 and 122.80 are obtained for the GHI and DNI respectively. Although, the clear-sky models, particularly the GHI, provide significantly better results than the simple linear regression, the model ignores critical variations in the day which makes it unreliable and unscalable for proper daily operation of the power system.

\subsection{Probabilistic Forecasting Method}
In order to demonstrate a probabilistic forecast method for this dataset, a quantile regression method was employed. This method may be considered more appropriate than a linear regression since conditions such as linearity and independence are not met in this case study. Also, while the linear regression method evaluates the mean value of a target variable, quantile regression helps to give information about the median of the target.

\begin{figure}[!t]
\centering
\includegraphics[width=0.51\textwidth]{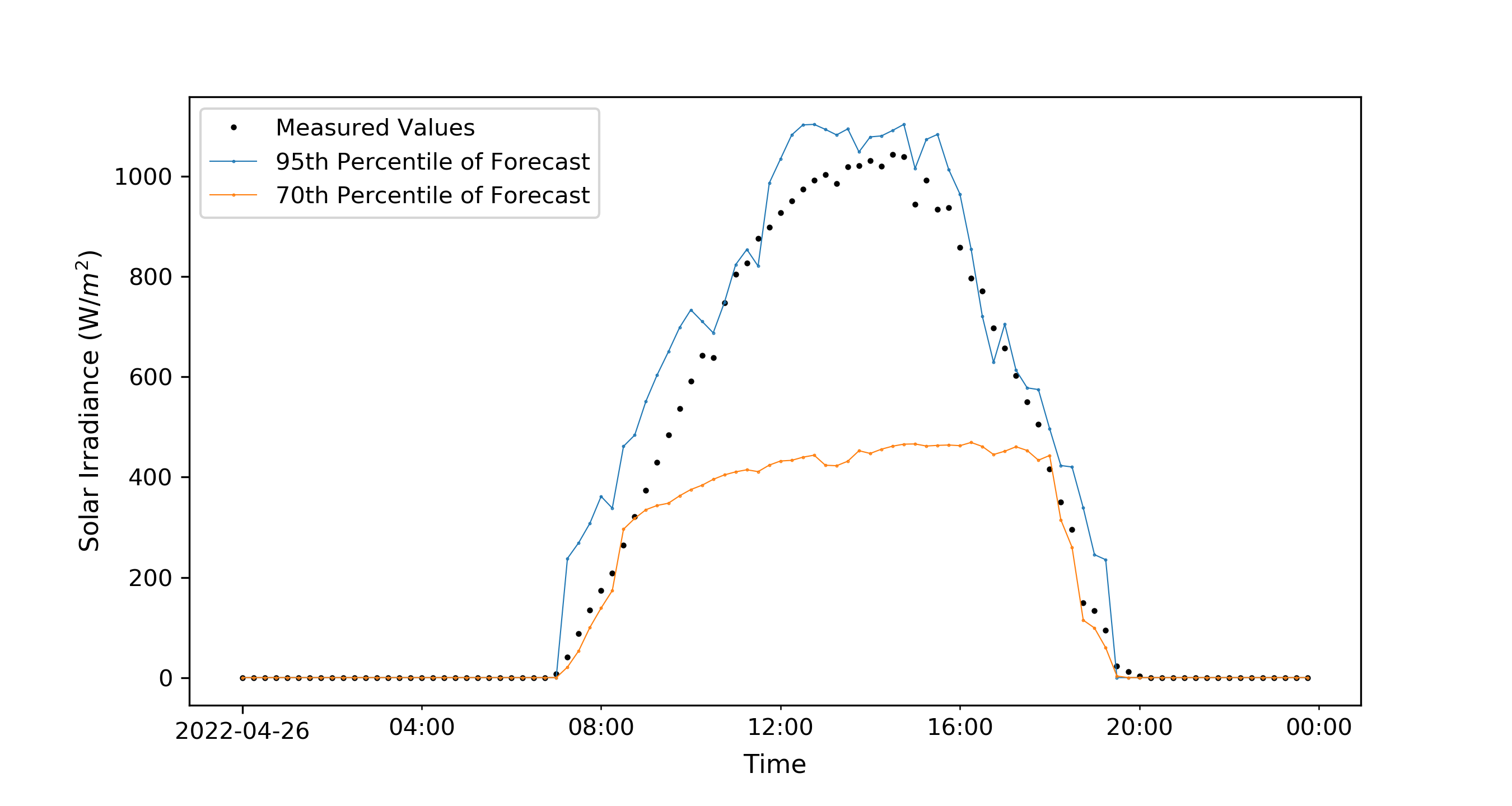}
\caption{Probabilistic forecast for 26th April (70th and 95th Percentile)}
\label{prob}
\end{figure}

For this case, the Temperature at 60cm measurement was used as independent variable to predict the solar irradiance. The same time period, as in the linear regression method, is used to train and test the model. Fig. \ref{prob} shows the probabilistic forecast generated using the quantile regression method. the results, specifically for the 70th and 95th percentiles, are illustrated. The different quantile range helps provide operators with additional information of the confidence of the prediction. The CRPS is used to assess the performance of this prediction and a value of 166.2 was obtained.

\subsection{Bayesian Forecasting Results and Discussion}
The Bayesian treatment discussed in Section III is applied to the data set to provide a probabilistic forecast of the solar irradiance. The same time period of the historic data as well, as the same independent variable used in the quantile regression method, are employed for the Bayesian approach. For this approach, the prior distribution of the parameters is assumed to be Gaussian with a mean (\(\mu_{0}\)) obtained from the historic data using weights obtained from a Gaussian copula method. The standard deviation \(\sigma_{0}\)) for this distribution is assumed to be 1. A Gaussian copula approach is used to determine the parameters for the prior distribution. Actual previous knowledge of the nature of the prior parameters improves the overall model by providing a distribution that is more symmetric and representative of the actual data.

\begin{figure}[!t]
\centering
\includegraphics[clip, trim=7cm 0.5cm 2cm 0.5cm, width=0.55\textwidth]{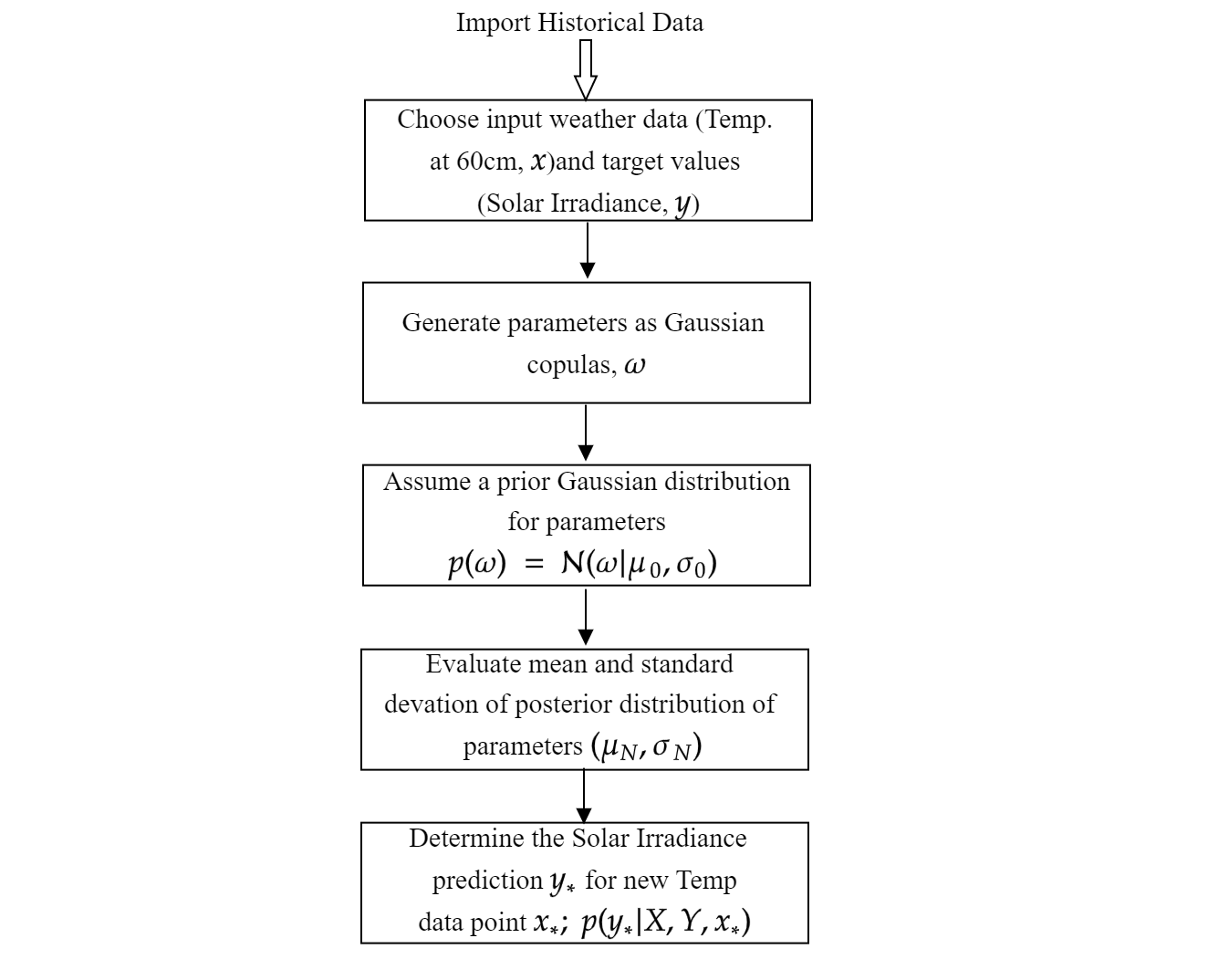} 
\caption{Bayesian Forecasting Algorithm}
\label{alg}
\end{figure}

The assumed prior Gaussian distribution is used in algorithms based on (\ref{posterior_mea}) and (\ref{posterior_sd}) to obtained the posterior distribution of the parameters. The prediction is then evaluated based on the parameters obtained. A summary of the algorithm developed for the the Bayesian method is presented in Fig. \ref{alg}

\begin{figure}[!t]
\centering
\includegraphics[width=0.52\textwidth]{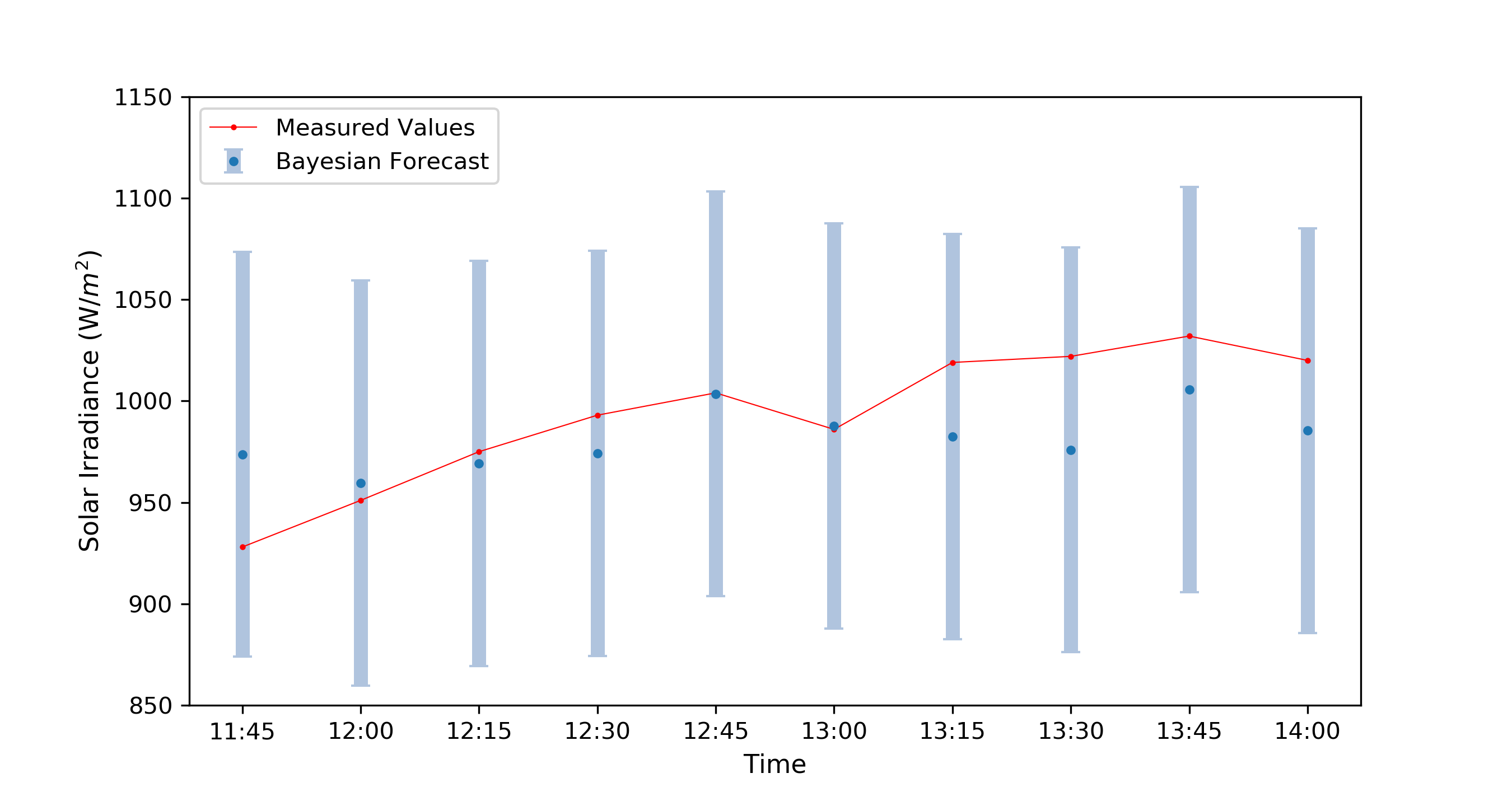}
\caption{Bayesian Forecast from 11:45am to 2:00pm}
\label{bayes}
\end{figure}

Fig. \ref{bayes} shows an error-bar plot based on the mean and standard deviation of the prediction for the time period 11:45am to 2:00pm. As can be seen from the illustration of the prediction for this period of the day, not only is the accuracy of the forecast improved but it provides a more informative approach for proper planning and operation purposes. The improvement in the prediction is revealed with a lower CRPS value of 27.72.

\begin{figure}[!t]
\centering
\includegraphics[width=0.52\textwidth]{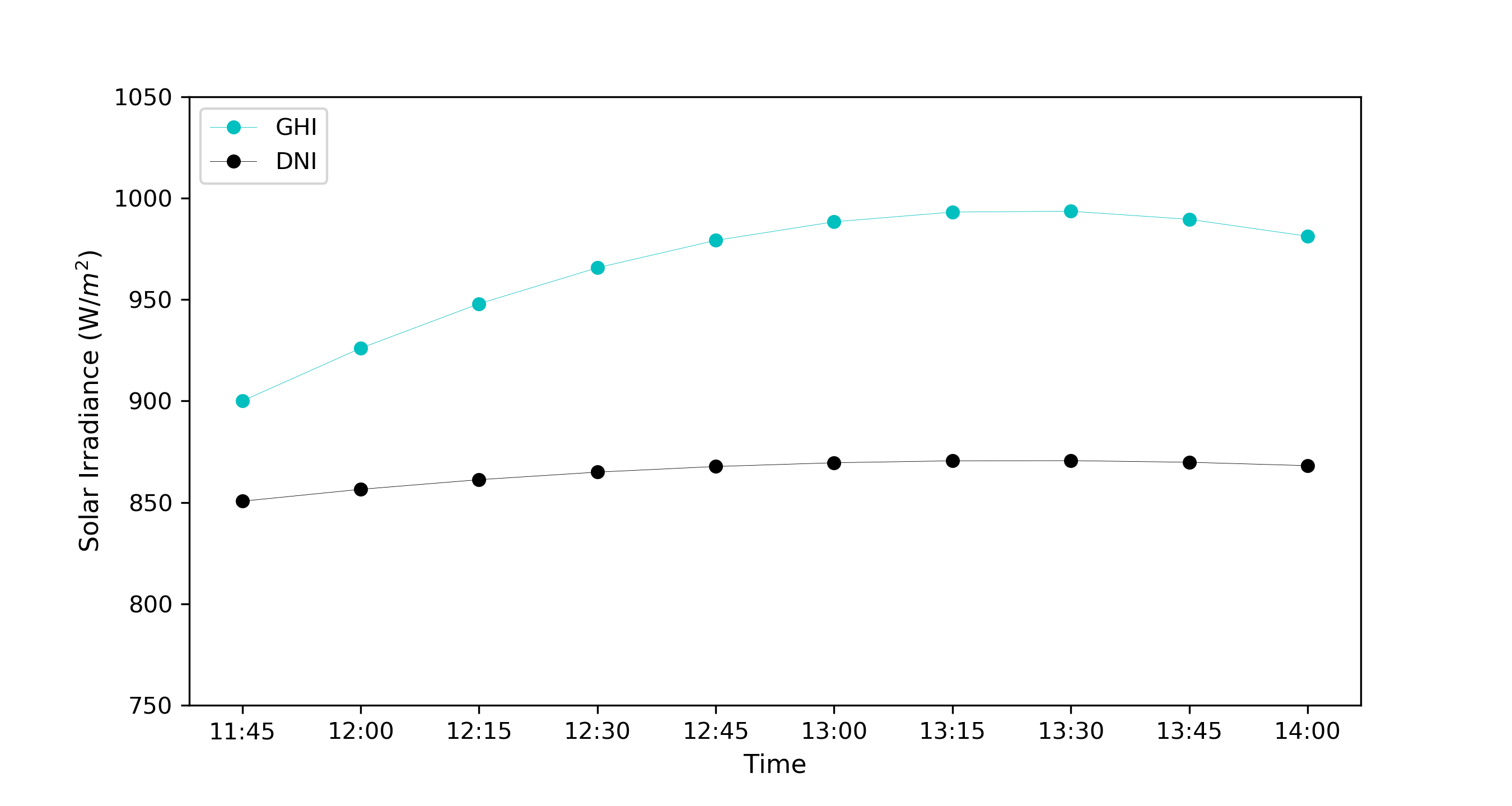}
\caption{GHI and DNI values from 11:45am to 2:00pm}
\label{GHI_times}
\end{figure}

Fig. \ref{GHI_times} shows the GHI and DNI values obtained from the clear-sky model, for the period between 11:45am and 2:00pm. The major advantage of the Bayesian method in comparison with the GHI is that, the probabilistic nature of the Bayesian method offers engineers critical margin for operational planning.

The Pinball Loss in (\ref{pinball}) is also used to evaluate the probabilistic forecasts. In this case study, \(q\) of 0.5, representing the 50th quantile is used for evaluation. This, in the case of the Bayesian method could be considered to be the mean value of the prediction. Table \ref{tab:Compare} provides the evaluation metrics for the quantile regression and Bayesian methods. It shows that the Bayesian approach provides significant improvement on the quantile regression method. A general improvement on the overall metrics can be obtained with the availability of data with stronger correlation to the solar irradiance.

\setlength{\arrayrulewidth}{0.01mm}
\begin{table}[ht]
\renewcommand\arraystretch{1.2}
\caption{Comparison of Quantile Regression (QR) and Bayesian Forecast Methods}
\label{tab:Compare}
\centering
\scalebox{1.2}{
\begin{tabular}{lll}
\hline 
\multicolumn{1}{c}{Method} & \multicolumn{1}{c}{Pinball Loss} & \multicolumn{1}{c}{CRPS} \\[0.1em]\hline
QR                           &88.17                                                                             &166.20                           \\[0.1em]
Bayesian                     & 52.72                                                                            &27.72                          \\[0.1em]\hline
\end{tabular}}
\end{table}

\section{Conclusion}
The modernization of the grid with a higher penetration of distributed renewable resources has necessitated the need for more comprehensive methods for power output forecasts. This is vital for grid operation. This paper discusses an approach to deal with the effects of intermittency and uncertainty in solar power output in the case of PV generation by providing a probabilistic forecast method for solar irradiance. The insufficiency of point forecast methods are discussed and demonstrated to show the value of probabilistic forecast methods.

A Bayesian treatment to solar irradiance forecasting is presented in this paper. It can be concluded that this method is risk-aware and more comprehensive in approach, by employing prior knowledge of the coefficients or parameters of the inputs used for prediction. Further research must focus on how to achieve symmetry between prior distributions assumed for the parameters and the actual characteristic distribution. Also for predictions to be accurate, researchers must identify appropriate input data that have stronger impact on solar irradiance.






\bibliographystyle{IEEEtran}

\bibliography{Reference}
%

\end{document}